\documentstyle[12pt]{article}

\begin{document}

\begin{flushright}
Preprint CAMTP/95-3\\
July 1995\\
\end{flushright}

\begin{center}
\large
{\bf Boundary integral method applied in chaotic quantum billiards}\\
\vspace{0.3in}
\normalsize
Baowen Li\footnote{e-mail Baowen.Li@UNI-MB.SI}
 and Marko Robnik\footnote{e-mail Robnik@UNI-MB.SI}\\

\vspace{0.2in}
Center for Applied Mathematics and Theoretical Physics,\\
University of Maribor, Krekova 2, SLO-62000 Maribor, Slovenia\\
\end{center}
\vspace{0.2in}
\normalsize
{\bf Abstract.}
The boundary integral method (BIM) is a formulation of Helmholtz equation in
the form of an integral equation suitable for numerical discretization to solve
the quantum billiard. This paper is an extensive numerical survey of  BIM in a
variety of quantum billiards, integrable (circle, rectangle), KAM  systems
(Robnik billiard) and fully chaotic (ergodic, such as stadium, Sinai billiard
and cardioid billiard). On the theoretical side we point out some serious flaws
in the derivation of BIM in the literature and show how the  final formula
(which nevertheless was correct) should be derived in a sound way and we also
argue that a simple minded application of BIM in nonconvex geometries presents
serious difficulties or even fails.  On the numerical side we have analyzed the
scaling of the averaged absolute  value of the systematic error $\Delta E$ of
the eigenenergy in units of mean level spacing with the density of
discretization ($b$ = number of numerical nodes on the boundary within one de
Broglie wavelength), and we find that in all cases the error obeys a power law
$ <|\Delta E|> = A b^{-\alpha}$, where $ \alpha $  (and also $A$) varies from
case to case (it is not universal), and is affected strongly  by the existence
of exterior chords in nonconvex geometries, whereas  the degree of the
classical chaos seems to be practically irrelevant.   We comment on the
semiclassical limit of BIM and make suggestions about a proper formulation with
correct semiclassical limit in nonconvex geometries.
\\\\

\\\\
Submitted to {\bf Journal of Physics A}

\normalsize
\vspace{0.3in}
\newpage

\section{Introduction}
In the studies of quantum chaos (Gutzwiller 1990, Giannoni {\em et al} 1989)
good numerical methods are not only indispensable but quite essential. We need
them in order to illustrate and verify the theoretical developments. Also, new
numerical experiments provide valuable material as evidence and inspiration for
new theories. Quantum billiards are certainly very useful model systems since
dynamically they are generic and - depending on the design - they can cover all
regimes of classical motion between integrability and full chaos (ergodicity).
Their classical dynamics can be easily followed for very long time periods
because no integration is necessary but only searching for zeros of certain
functions at collision points. Quantanlly they have the advantage of having a
compact configuration space admitting application of various numerical methods
such as the Boundary Integral Method (BIM) (Banerjee 1994, Berry and Wilkinson
1984, henceforth referred to as BW,  and the references therein), the Plane
Wave Decomposition Method (PWDM) employed by Heller (1984), conformal mapping
diagonalization technique introduced by Robnik (1984) and further developed by
Berry and  Robnik (1986) and recently by Prosen and Robnik (1993, 1994), also
by  (Bohigas {\em et al} 1993),  and a number of other methods, which, however,
might be adapted to some special systems.  Among the general methods BIM is
probably most widely used, even in quite practical engineering problems. The
main task of the present paper is to point out some logical flaws in deriving
this method in the context of the existing literature, to analyze its
limitations in cases of nonconvex geometries with suggestions for improvements
and generalizations, and to perform an extensive numerical investigation of its
accuracy in relation to geometrical properties and classical dynamics.
\\\\
In order to clearly expose the  difficulties and the errors in the derivation
of BIM offered in the literature, e.g.  in BW, we present our {\em regularized}
derivation, by which we mean that we construct and use a Green function which
automatically (by construction) satisfies the Dirichlet boundary condition
(vanishes on the boundary $\partial {\cal B}$ of the billiard domain ${\cal
B}$), which is achieved by employing the method of images (see e.g. Balian and
Bloch (1974) and the references therein).  This will enable us to avoid
committing two errors, which, however, luckily compensated each other: {\em
Firstly}, in taking the normal derivatives on the  two sides of equation (6) in
BW, on the rhs we must use the value of $\psi({\bf r})$ which is the interior
solution inside ${\cal B}$ rather than $\frac{1}{2}\psi({\bf r})$ which is the
value exactly on the boundary, simply because in taking the derivatives we must
evaluate the function at two infinitesimally separated points normal to the
boundary; {\em Secondly}, this error of taking the unjustified factor $1/2$ is
then exactly compensated by another error in arriving at the equation (8) in
BW, namely by interchanging the integration along the boundary $\partial {\cal
B}$  and the normal differentiation, because due to singularities on the
boundary  these two operations do not commute.
\\\\
So, let us offer our regularized derivation. We are searching for the solution
$\psi({\bf r})$ with eigenenergy $E = k^2$ obeying the Helmholtz equation
\begin{equation}
\nabla_{{\bf r}}^2 \psi({\bf r}) + k^2 \psi({\bf r}) = 0,
\label{eq:Helmholtz}
\end{equation}
with the Dirichlet boundary condition $\psi({\bf r})=0$ on the boundary
 ${\bf r} \in \partial {\cal B}$.
We will transform this Schr\"odinger equation for our quantum billiard ${\cal
B}$ into an integral equation by means of the {\em regularized} Green function
$G({\bf r}, {\bf r}^{\prime})$, which solves the following defining equation
\begin{equation}
\nabla_{{\bf r}}^2 G({\bf r},{\bf r'})  + k^2 G({\bf r},{\bf r'})
= \delta({\bf r}-{\bf r'}) - \delta({\bf r}-{\bf r'}_{\it R}),
\label{eq:Green-eq}
\end{equation}
where ${\bf r}$ and  ${\bf r'}$ are in $ {\cal B}\cup {\partial {\cal B}}$,
and ${\bf r'}_{\it R}$ is the mirror image of ${\bf r'}$ with respect to the
tangent at the closest lying point on the boundary, and thus if ${\bf r'}$
is sufficiently close to the boundary then ${\bf r'}_{\it R}$ is outside the
billiard ${\cal B}$.
The solution can easily be found in terms of the free propagator (the free
particle Green function on the full Euclidean plane)
\begin{equation}
G_0({\bf r}, {\bf r'}) = - \frac{1}{4} i H_{0}^{(1)}(k|{\bf r}-{\bf r'}|),
\label{eq:Gfree}
\end{equation}
where $H_{0}^{(1)}$ is the zero order Hankel function of the first kind
(Abramowitz and Stegun 1972), namely
\begin{equation}
G({\bf r},{\bf r'}) = G_{0}({\bf r},{\bf r'}) -G_{0}({\bf r},{\bf r'}_{\it R}),
\label{eq:Green}
\end{equation}
such that now $G({\bf r},{\bf r'})$ is zero by construction for any
${\bf r'}$ on the boundary $\partial {\cal B}$, in contradistinction to the
Green function defined and used in equation (5) in BW.
Multiplication of the equation (\ref{eq:Green-eq}) by $\psi({\bf r})$
and the Helmholtz equation (\ref{eq:Helmholtz}) by $G({\bf r}, {\bf r'})$,
subtraction, integration over the area inside ${\cal B}$ and using Green's
theorem, yields
\begin{equation}
\oint ds\left(\psi({\bf r}){\bf n}\cdot\nabla_{{\bf r}} G({\bf r},{\bf r'})
-G({\bf r},{\bf r'}){\bf n}\cdot\nabla_{{\bf r}} \psi({\bf r}) \right)
= \psi({\bf r'}),
\label{eq:BIM1}
\end{equation}
where $s$ is the arclength on the boundary $\partial {\cal B}$ oriented
anticlockwise, ${\bf n}$ is the unit normal vector to $\partial {\cal B}$
at ${\bf r}$
oriented outward, and this equation is now valid for {\em all} ${\bf r'}$
inside
and on the boundary of ${\cal B}$. Since in this equation everything is
regular we
can take the normal partial derivatives on both sides. Following the usual
notation in BW we define the normal derivative of $\psi$ at the point $s$ as
\begin{equation}
u(s) = {\bf n}\cdot\nabla_{{\bf r}}\psi({\bf r}(s)),
\label{eq:nder}
\end{equation}
and thus using the boundary condition $\psi({\bf r}) =0$ we arrive at
\begin{equation}
u(s) = -2\oint ds'u(s'){\bf n}\cdot\nabla_{{\bf r}}G_{0}({\bf r},{\bf r'}),
\label{eq:BIM2}
\end{equation}
In this way we have correctly derived the main integral equation of the
boundary integral method which is correctly given as equation (8) in BW (where
the two errors exactly compensate), so that all the further steps in working
out
the geometry of equation (\ref{eq:BIM2}) and the numerical discretization are
exactly the same as in BW. As shown in figure 1 we define the length of the
chord between two points on the boundary ${\bf r}(s)$ and ${\bf r}(s')$ as
\begin{equation}
\rho(s,s') = |{\bf r}(s) -{\bf r}(s')|
\label{eq:chord}
\end{equation}
and the angle $\theta(s',s)$ is the angle between the chord and the tangent to
$\partial {\cal B}$ at $s'$. Of course $\theta(s,s') \neq \theta (s',s)$.
Thus following the notation of BW we can write
\begin{equation}
{\bf n'}\cdot\frac{\partial G_0({\bf r},{\bf r'})}{\partial {\bf r'}}
= \sin\theta(s',s)\frac{\partial G_0}{\partial \rho}
\label{eq:normgreen}
\end{equation}
and we obtain finally
\begin{equation}
u(s) = -\frac{1}{2} ik\oint ds' u(s')\sin\theta(s,s')H_1^{(1)}\{k\rho(s,s')\}.
\label{eq:bim3}
\end{equation}
In numerically solving this integral equation we have used precisely the same
primative discretization procedure as BW, which turned out to be better than
some other more sophisticated versions. So we simply divide the perimeter
${\cal L}$ into $N$ equally long segments and thus define
\begin{equation}
s_m = m{\cal L}/N,\quad \rho(s_l,s_m) =\rho_{lm},\quad \theta(s_l,s_m)=
\theta_{lm},\quad l\le l,m \le N,
\label{eq:notation}
\end{equation}
Therefore numerically we are searching for the zero of the determinant
$\Delta_N(E) = det(M_{lm})$ where $M_{lm}$ are the matrix elements
of the $N\times N$ matrix,
\begin{equation}
M_{lm} = \delta_{lm} + \frac{ik{\cal L}}{2N} \sin\theta_{lm}
H_1^{(1)}(k\rho_{lm}),
\label{eq:Mmatrix}
\end{equation}
where $E=k^2$. Due to the asymmetry $\theta_{lm} \neq \theta_{ml}$ this matrix
is a general complex non-Hermitian matrix.
\\\\
One important aspect of this formalism is the semiclassical limiting form which
has been extensively studied by Boasman (1994) and which is the subject of our
current investigation (Li and Robnik 1995a). At this point we just want to make
the following comment. In cases of nonconvex geometries we will have {\em
exterior} chords connecting two points on the boundary such that they lie
entirely, or at least partially,  outside ${\cal B}$. While formally the method
and the procedure in such cases is perfectly right, in reality it fails
completely, and this can be seen by considering the semiclassical limit. The
formal leading order in the asymtotic expansion of the Hankel function
$H_1^{(1)}$ (Debye approximation) does not match the actually correct
semiclassical leading approximation which would be spanned by the shortest
classical orbit connecting the two points via at least one or many collision
points. Therefore we understand and expect that the method must fail or at
least must meet severe difficulties in cases of nonconvex geometry. This has
been completely confirmed in our present work as we will show in the next
section, whilst the analytical work to
reformulate the method including the multiple collision expansions is in
progress (Li and Robnik 1995b) and is expected to deal satisfactorily with
nonconvex geometries.
\\\\
In the next section we shall analyze the numerical accuracy of BIM as a
function of the density of discretization
\begin{equation}
b = \frac{2\pi N}{k{\cal L}},
\label{eq:defb}
\end{equation}
in a variety of quantum billiards with integrable, KAM-type or ergodic
classical dynamics, including such with nonconvex geometry. The main result is
that there is always a power law so that the error of eigenenergy in units of
mean level spacing, after taking average of the absolute value over a suitable
ensemble of eigenstates,  obeys $<|\Delta E|> = A b^{-\alpha}$,
but the exponent $\alpha$ is
nonuniversal and typically becomes almost zero if there are nonconvex segments
on the boundary.
\\\\
\section{Numerical results}

The numerical procedure we have used to solve the variety of quantum billiards
is exactly as described in the introduction and therefore it is precisely the
same as in BW. Our main task in this work is to analyze in detail the behavior
of BIM as a function of the density of discretization $b$, especially in
relation to the geometrical properties of ${\cal B}$ (nonconvexities) and in
relation to classical dynamics, whose chaotic behaviour is expected to imply
interesting methodological and algorithmical manifestation of quantum chaos.
\\\\
This to end we have to measure the numerical error of the eigenenergies in some
natural units, which obviously is the mean level spacing. In plane billiards
this is well defined and determined by the leading term of the Weyl formula,
namely in our units it is equal to $4\pi/{\cal A}$, where ${\cal A}$ is
the area of the billiard ${\cal B}$. Since the error $\Delta E$
of the eigenenergies thus
defined still fluctuates widely from state to state we have to perform some
kind of averaging over a suitable ensemble of  consecutive states. But this
will make sense only if such a local average of the error is stationary
(constant) over a suitable energy interval. This condition has been confirmed
to be satisfied  in almost all cases which we checked. In case of the circle
billiard and in case of half circle billiard (which embodies all the odd
states of the full  circle billiard) this stationarity of the locally averaged
error is shown in figures 2(a-b), respectively.
\\\\
Having established that we have always taken the average of the absolute value
of the error (in units of the mean level spacing)
over a suitable ensemble of eigenstates, for which we have chosen
the lowest one hundred eigenstates in all cases. This quantity will be
henceforth denoted by $<|\Delta E|>$.
\\\\
In figure 3(a-f) we show the error $<|\Delta E|>$
versus $b$ for three shapes of
the Robnik billiard with the shape parameter $\lambda= 0, 1/4, 1/2$. (The
billiard shape is defined by the quadratic conformal map of the unit disk
$|z|\le 1$ onto to the $w$-plane, namely $w(z) = z +\lambda z^2$.) We show the
normal plot of the averaged absolute value of the error versus $b$ in the
figures 3(a,c,e) and the log-log plot in figures 3(b,d,f).
The best fitting power law curve is described by
\begin{equation}
<|\Delta E|> = A b^{-\alpha},
\label{eq:power-law}
\end{equation}
and is seen to provide a very significant fit in all three cases.
One should be reminded that at $\lambda =0$ we have integrable classical
dynamics in the circle billiard, at $\lambda =1/4$ we have almost ergodic but
nevertheless KAM-type dynamics with very tiny islands of stability
(see e.g. Li and Robnik (1994,1995c); it should
be emphasized that at $\lambda=1/4$ we have zero curvature point at $z=-1$, and
for all $\lambda > 1/4$ the shape is nonconvex),
whilst at
$\lambda = 1/2$ we have rigorous ergodicity (Markarian 1993) and also nonconvex
geometry. So one can observe that increasing chaos from integrability (circle)
to almost ergodicity ($\lambda =1/4$) has almost no effect on $\alpha$, whereas
the nonconvexities at $\lambda > 1/4$ seem to imply a complete "collapse" of
$\alpha$: As it will be shown in figure 5(a) $\alpha$ drops to zero almost
discontinuously at $\lambda =1/4$ and then increases up to $0.4$ at $\lambda$
close to $1/2$.  This is because, perhaps, close to $\lambda =1/2$ where we
have
the cusp singularity at $z =-1$, the role of the nonconvexities is smaller.
In all cases we have calculated all states by applying BIM but then for
technical reasons compared only the odd states with their exact value, which
are supplied by the conformal mapping diagonalization technique (see e.g.
Robnik 1984, Prosen and Robnik 1993).
\\\\
It is then interesting to look at the similar plots for BIM as applied to half
billiard by which we mean the upper part of the Robnik billiard defined for
$\Im(z)\ge 0$, implying also $\Im (w)\ge 0$, still with the Dirichlet boundary
conditions everywhere on the
boundary.  This embodies all the odd states of the full billiard.
The classical dynamics in the two billiards is of course exactly the same and
yet in figures 4(a-f) for the half billiard we see that $\alpha$ is now
notably different, showing that there is no universality in the value of
$\alpha$ so that $\alpha$ is certainly not uniquely determined by the classical
dynamics of the underlying quantum billiard. On the other hand we can also see
the effect of nonconvexities of the boundary $\partial{\cal B}$: Namely for the
half (cardioid) billiard  at $\lambda =1/2$ we have {\em no} nonconvexities and
this probably is precisely the reason for large increase of $\alpha$ from $0.4$
in figure 3(e-f) to $2.4$ in figure 4(e-f).
\\\\
Having established the validity of the power law (\ref{eq:power-law}) it is now
most interesting and also immensely CPU-time consuming  (it took almost one
month of CPU-time on Convex C3860 to produce figure 5(a,c) and a little bit
less for figure 5(b,d)) to look at the variation of $\alpha$ with the billiard
shape parameter $\lambda$. For the full Robnik billiard this is shown in figure
5(a) and for the half billiard in figure 5(b). In both cases there is a flat
region of almost constant $\alpha$ within $0 \le \lambda \le 1/4$: In the
former case it fluctuates slightly around $3.5$ and in the latter case it is
surprisingly stable around $2.9$. At $\lambda > 1/4$ the nonconvexities of
boundary appear and this implies - as explained qualitatively in the
introduction - strong drop of $\alpha$ which is much sharper (almost
discontinuous) in case of the full billiard because obviously the
nonconvexities
are more pronounced there than in the half billiard. So in the former case
$\alpha$ drops almost to zero, whereas in the half billiard it decreases rather
smoothly to about $0.16$ at $\lambda =0.36$, and then increases again reaching
the value of $2.4$ at $\lambda =1/2$.
\\\\
Apart from $\alpha$ in (\ref{eq:power-law}) we would also like to know the
value of the constant $A$ (the pre-factor) in each case. This is given by
fixing $b=12$ and plotting the mean absolute value of the error
(averaged over the lowest one hundred
odd states) versus $\lambda$ logarithmically. Here we see that in both cases
the mean error $<|\Delta E|>$ is almost constant up to $\lambda \le 1/4$ and is
about $5\times 10^{-6}$ for the full billiard whilst for the half billiard
it is about $7\times 10^{-5}$. This discrepancy by almost an order of magnitude
is not completely understood. At $\lambda \ge 1/4$ please observe the dramatic
increasing of $<|\Delta E|>$. For the full billiard  it increases almost
discontinuously up to $2\times 10^{-2}$ whereas in the half billiard it reaches
the minimum of $4\times 10^{-3}$ at about $\lambda =0.36$ and decreases then
again to $2\times 10^{-4}$ at $\lambda =1/2$. This difference again is
qualitatively explained by the role of nonconvexities.
\\\\
Finally in figure 6(a-f) we show the worst cases of applying BIM, namely the
Robnik billiard at $\lambda=0.4$ in figures 6(a,b), the half billiard in
figures 6(c,d) with the same value of $\lambda=0.4$ and the desymmetrized Sinai
billiard (1/8 of the full billiard) with the unit side length and radius
$R=1/4$ of the circular obstacle. In all cases $\alpha$ is almost zero  or
even slightly negative which implies that there is practically no convergence
of the numerical eigenvalues by increasing the density of discretization $b$.
Clearly, this is due to the nonconvex geometry of these billiards.
\\\\
Most of our results are summarized in table 1 for three classes of billiard
systems with different type of classical dynamics, namely integrable, KAM-type
and ergodic systems. We show the calculated values of $\alpha$ and also the
average absolute value of the error $<|\Delta E|>$ with fixed value of $b=12$.
The table clearly demonstrates that the power law (\ref{eq:power-law})
for BIM is universal but not the exponent $\alpha$ and the prefactor $A$.
It also demonstrates that
classical dynamics has little effect on $\alpha$ whereas the nonconvexities of
the boundary are quite crucial: In all cases of pronounced nonconvex
geometry $\alpha$ is
typically close to zero or even slightly negative like in the Sinai billiard.
In the table we include also the results on the integrable case of the
rectangular equilateral triangle (half of the unit square) where
$\alpha=3.28$ is quite large, and the ergodic case of the $1/4$ Heller's
stadium
($2\times 2$ square plus two semicircles with unit radius) in which case
$\alpha =3.0$ is also quite large. Both of these two cases have convex
geometry and thus large $\alpha$ despite the completely different classical
dynamics.
\\\\
When thinking about improving the efficiency and the accuracy of BIM we have
also tried a more sophisticated version of BIM, where we have explicitly used
a Gaussian integration on the boundary when discretizing our main equation
(\ref{eq:BIM2}). However, this experience has been negative after many carefull
checks in various billiards, and therefore we decided to resort to the
primitive discretization of BIM which is exactly the same approach as in BW.

\section{Discussion and conclusions}
The main purpose of this paper is to investigate the behaviour of the Boundary
Integral Method (BIM) with respect to the density of discretization $b$ as
defined in equation (\ref{eq:defb}) ($b$ = the number of numerical nodes per de
Broglie wavelength along the boundary).
In all cases we discovered that there is a power law behaviour described in
(\ref{eq:power-law}): The average absolute value of the error measured in the
units of mean level spacing is given by $<|\Delta E|> = A b^{-\alpha}$.
We wanted to verify whether there is
any systematic effect of classical dynamics of quantum billiards on $\alpha$
and $A$. The answer is negative. On the other hand we found that the role of
nonconvex geometry of the boundary is crucial: If there is a nonconvex part of
the boundary $\alpha$ is typically close to zero or even slightly negative
implying that there is practically no convergence of the eigenvalues with
respect to the increasing $b$. This failure of BIM can be understood as
explained in the introduction, and the easiest way to see that is to consider
the semiclassical limiting approximation of BIM (Li and Robnik 1995a).
After explaining two
systematic errors in the literature where the integral BIM equation is derived
and where luckily the two errors mutually exactly compensate, we have given
the correct (regularized) derivation and discussed
the BIM formalism thus derived. We agree that even in nonconvex geometries it
is formally right, but nevertheless practically fails which is most clearly
demonstrated by the semiclassical limit mentioned above.
Therefore we suggest a generalization
of the BIM method by using multiple reflection (collision) expansion which is
another subject of our current investigation (Li and Robnik 1995b), and is
important not only for studies in quantum chaos but also in engineering
problems (Banerjee 1994).
\\\\
Most of our results are summarized in table 1, giving the evidence for the
above conclusions. In another work (Li and Robnik 1995d) we have investigated
the impact of classical chaos on another general numerical method for quantum
billiards, namely the plane wave decomposition method employed by Heller (1984,
1991), where we have found that at fixed $b$ the average absolute value of the
error $<|\Delta E|>$ does correlate with classical chaos and increases sharply
with increasing classical chaos. In this method such a behaviour can be
understood much more easily: In classically integrable cases the wavefunction
in the semiclassical limit can be correctly described locally by a finite
number of plane waves, whereas in classically fully chaotic (ergodic) systems
we need locally infinite number of plane waves. Therefore at fixed $b$ the
accuracy of Heller's method strongly deteriorates as the system approaches
ergodicity. More results on that will be published in a separate paper (Li and
Robnik 1995d).
\\\\
It remains an interesting and important theoretical problem to study the
sensitivity of the eigenstates (eigenenergies and wavefunctions) on the
boundary data of eigenfunctions, of which one aspect is also the dependence of
the eigenstates on the billiard shape parameter. If such sensitivity correlates
with classical chaotic dynamics and at the same time manifests itself in the
accuracy of the purely quantal numerical methods then such a behaviour would be
one important manifestation of quantum chaos. This interesting line of thoughts
in the search of another face of quantum chaos will be further developed in
another work (Li and Robnik 1995e) where we also present detailed studies of
level curvature distribution  and other measures of the sensitivity of the
eigenstates.

\section*{Acknowledgments}
We thank Dr. Holger Schanz and Professor Uzy Smilansky for the table of the
eigenenergies for the Sinai billiard, and Dr. Toma\v z Prosen for
supplying the eigenenergies of Robnik billiard.
The financial support by the Ministry of Science and
Technology of the Republic of Slovenia is gratefully acknowledged.

\vfill
\newpage
\section*{References}

Abramowitz, M and Stegun I A (eds) 1972 {\em Handbook of mathematical
functions} Dover, New York\\\\
Balian R and Bloch C 1974 {\em Ann. Phys} {\bf 85} 514\\\\
Banerjee P K 1994 {\em The Boundary Element Methods in Engineering},
McGraw-Hill Book Company. London\\\\
Berry M V and Robnik M 1986 {\em J. Phys. A: Math. Gen.} {\bf 19} 649\\\\
Berry M V and Wilkinson M 1984 {\em Proc. R. Soc. Lond. A} {\bf 392} 15\\\\
Boasman P A 1994 {\em Nonlinearity} {\bf 7} 485\\\\
Bohigas O, Boos\'e D, Egydio de Carvalho R and Marvulle V 1993 {\em
Nuclear Physics A}{\bf 560} 197\\\\
Giannoni M-J, Voros J and Zinn-Justin eds. 1991 {\em Chaos and Quantum Systems}
(North-Holland)\\\\
Gutzwiller M C 1990 {\em Chaos in Classical and Quantum Mechanics} (New York:
Springer)\\\\
Heller E J 1984 {\em Phys. Rev. Lett} {\bf 53} 1515\\\\
Heller E J 1991  in {\em Chaos and Quantum Systems (Proc. NATO ASI Les Houches
Summer School)} eds M-J Giannoni, A Voros and J Zinn-Justin,
(Amsterdam: Elsevier) p547\\\\
Li Baowen and Robnik M 1994 {\em J. Phys. A: Math. Gen.} {\bf 27} 5509\\\\
Li Baowen and Robnik M 1995a to be submitted to {\em J. Phys. A: Math.
Gen.}\\\\
Li Baowen and Robnik M 1995b to be submitted to {\em J. Phys. A: Math.
Gen.}\\\\
Li Baowen and Robnik M 1995c {\em J. Phys. A: Math. Gen.} {\bf 28} 2799\\\\
Li Baowen and Robnik M 1995d to be submitted to {\em J. Phys. A: Math.
Gen.}\\\\
Li Baowen and Robnik M 1995e to be submitted to {\em J. Phys. A: Math.
Gen.}\\\\
Markarian R 1993 {\em Nonlinearity} {\bf 6} 819\\\\
Prosen T and Robnik M 1993 {\em J. Phys. A: Math. Gen.} {\bf 26} 2371\\\\
Prosen T and Robnik M 1994  {\em J. Phys. A: Math. Gen.} {\bf 27} 8059\\\\
Robnik M 1984 {\em J. Phys. A: Math. Gen.} {\bf 17} 1049\\\\

\newpage
\section*{Table}

\bigskip
\bigskip

 {\bf Table 1.} The power law exponent $\alpha$  and the average absolute value
of the error $<|\Delta E|>$ with $b=12$ for different billiards. For details of
the KAM-type see also figures 5(a,b).
\vspace{15mm}\\

\begin{tabular}{|l|l|c|c|}\hline
& & \\
Type & Quantum billiard  & $\alpha$ & $<|\Delta E|>_{b=12}$ \\[2.5ex] \hline
& & & \\
  & Circle (half)  & 2.94 $\pm$ 0.17 & 6.74E$-$5\\[1ex]
Integrable & Circle (full)  & 3.44 $\pm$ 0.18 & 5.97E$-$6\\[1ex]
& Rectangle-triangle & 3.28 $\pm$ 0.29 & 4.08E$-$5\\[1ex]\hline
& & &  \\
& Robnik (full) & & \\ [-.3ex]
KAM &($0< \lambda < 1/4)$ & $\approx 3.4$ & $\approx$ 5.0E$-$6\\[1.2ex]
& Robnik (half) & & \\ [-.3ex]
& ($0< \lambda < 1/4$) & $\approx$ 2.9 & $\approx$ 7.0E$-$5\\[1ex]
\hline
& & &\\
&  Stadium (1/4) &   3.00 $\pm$ 0.16 & 1.18E$-$4\\[1ex]
 & Cardioid (half) & 2.42 $\pm$ 0.11 &  1.76E$-$4\\[1ex]
 & Cardioid (full) & 0.42 $\pm$ 0.08 & 1.85E$-$2\\[1ex]
Ergodic& Sinai (1/8)  & $-$ 0.34 $\pm$ 0.11 & 3.63E$-$1\\[1ex]
& Robnik (full) &  & \\[-0.3ex]
& ($0.3 < \lambda < 1/2$) & see figure 5a & see figure 5c\\[1.2ex]
& Robnik (half) &  & \\[-0.3ex]
& ($ 0.3 < \lambda < 1/2) $ &see figure 5b & see figure 5d\\[1ex]
 \hline
\end{tabular}

\vfill

\newpage
\section*{Figure captions}

\bigskip
\bigskip

\noindent {\bf Figure 1:}
The notation of the angles and chords used in BIM.
\bigskip
\bigskip

\noindent {\bf Figure 2:}
The BIM error (measured in units of mean level spacing) of eigenstates
versus energy. The error is difference between
the BIM value and the exact value. The lowest 1000 odd states are shown. Plot
(a) is for the full circle billiard and (b) for the half circle billiard. In
both cases $b$ is fixed, $b=6$.
\bigskip
\bigskip

\noindent {\bf Figure 3:}
The ensemble averaged (over 100 lowest odd eigenstates) absolute BIM error
versus the density of boundary discretization $b$ and the best power law
fit for the full Robnik billiard at different shape parameters $\lambda$.
The $+$ represents the numerical data, the curve is the best power law fit
whose $\alpha$ is given in (b,d,f). Figures (b,d,f) are log-log plots of the
same quantaties as in (a,c,e).
\bigskip
\bigskip

\noindent {\bf Figure 4:}
The same as in figure 3 but for the half Robnik billiard.
\bigskip
\bigskip

\noindent {\bf Figure 5:}
We show  $\alpha$ versus $\lambda$ plot for Robnik billiard in (a) and for half
Robnik billiard in (b). In (c) and (d) we plot $-\lg(<|\Delta E|>)$ with fixed
$b=12$ versus $\lambda$ for the two billiards, respectively.

\bigskip
\bigskip
\noindent {\bf Figure 6:}
We show three examples  of worst cases in applying BIM, Robnik billiard with
$\lambda=0.4$ in (a,b), half Robnik billiard with the same $\lambda$ in (c,d),
and desymmetrized  Sinai billiard ($1/8$) with unit side length and radius
$R=1/4$. In (a,c,e) we plot the averaged absolute value of the error versus
$b$, and in (b,d,f) we plot log-log diagrams of the same quantities.

\end{document}